\documentclass[11pt]{article}
\usepackage[pctex32]{graphics}

\def\ben{\begin{equation}}
\def\een{\end{equation}}

\def\nn{\nonumber} \def\bd{\begin{document}} \def\ed{\end{document}}
\def\ds{\documentstyle} \let\fr=\frac \let\bl=\bigl \let\br=\bigr
\let\Br=\Bigr \let\Bl=\Bigl
\let\bm=\bibitem
\let\na=\nabla
\let\pa=\partial \let\ov=\overline
\newcommand{\be}{\begin{equation}}
\newcommand{\ee}{\end{equation}}
\def\ba{\begin{array}}
\def\ea{\end{array}}
\def\ft#1#2{{\textstyle{\frac{\scriptstyle #1}{\scriptstyle #2} } }}
\def\fft#1#2{{\frac{#1}{#2}}}
\def\del{\partial}
\def\vp{\varphi}
\def\sst#1{{\scriptscriptstyle #1}}
\def\oneone{\rlap 1\mkern4mu{\rm l}}
\def\td{\tilde}
\def\wtd{\widetilde}
\def\ie{{\it i.e.\ }}
\def\dalemb#1#2{{\vbox{\hrule height .#2pt
        \hbox{\vrule width.#2pt height#1pt \kern#1pt
                \vrule width.#2pt}
        \hrule height.#2pt}}}
\def\square{\mathord{\dalemb{6.8}{7}\hbox{\hskip1pt}}}
\newcommand{\ho}[1]{$\, ^{#1}$}
\newcommand{\hoch}[1]{$\, ^{#1}$}
\newcommand{\bea}{\setlength\arraycolsep{2pt} \begin{eqnarray}}
\newcommand{\eea}{\end{eqnarray}}
\newcommand{\ra}{\rightarrow}
\newcommand{\lra}{\longrightarrow}
\newcommand{\Lra}{\Leftrightarrow}
\newcommand{\bp}{\tilde \beta^\prime}
\newcommand{\tr}{{\rm tr} }
\newcommand{\Tr}{{\rm Tr} }
\def\0{{\sst{(0)}}}
\def\1{{\sst{(1)}}}
\def\2{{\sst{(2)}}}
\def\3{{\sst{(3)}}}
\def\4{{\sst{(4)}}}
\def\5{{\sst{(5)}}}
\def\6{{\sst{(6)}}}
\def\7{{\sst{(7)}}}
\def\8{{\sst{(8)}}}
\def\m{{\sst{(m)}}}
\def\n{{\sst{(n)}}}
\def\cA{{{\cal A}}}
\def\cB{{{\cal B}}}
\def\cF{{{\cal F}}}
\def\cG{{{\cal G}}}
\def\cH{{{\cal H}}}
\def\tV{\widetilde V}
\def\tW{\widetilde W}
\def\tH{\widetilde H}
\def\tE{\widetilde E}
\def\tF{\widetilde F}
\def\tA{\widetilde A}
\def\im{{{\rm i}}}
\def\tY{{{\wtd Y}}}
\def\ep{{\epsilon}}
\def\vep{{\varepsilon}}
\def\bD{{{\bar D}}}
\def\R{{{\mathbb R}}}
\def\C{{{\mathbb C}}}
\def\H{{{\mathbb H}}}
\def\CP{{{\mathbb C}{\mathbb P}}}
\def\RP{{{\mathbb R}{\mathbb P}}}
\def\Z{{{\mathbb Z}}}
\def\bA{{{\mathbb A}}}
\def\bB{{{\mathbb B}}}
\def\bC{{{\mathbb C}}}
\def\bD{{{\mathbb D}}}
\def\bE{{{\mathbb E}}}
\def\bZ{{{\mathbb Z}}}
\def\Re{{{\frak{Re}}}}
\def\Im{{{\frak{Im}}}}
\def\cosec{{\,\hbox{cosec}\,}}
\def\Gm{{\Gamma_{\!\! -}}}
\def\Gp{{\Gamma_{\!\! +}}}
\def\stan{{standard }}
\def\nonstan{{supernumerary }}
\def\p{{\partial}}
\def\kdel#1{{\fft{\del}{\del#1}}}

\def\bog{{Bogomolny }}
\def\om{{\omega}}

\newcommand{\nnr}{\nonumber \\}
\newcommand{\pd}{\partial}
\newcommand{\ud}{\textrm{d}}
\newcommand{\dTH}{T^{\prime \, 0}_\textrm{H}}
\newcommand{\dOi}{\Omega^{\prime \, 0}_i}
\newcommand{\bx}{{\bf x}}

\begin{document}

\vspace{5mm}
\begin{center}
{\Large \bf Nonpropagation of scalar in  the deformed
Ho\v{r}ava-Lifshitz gravity } \vspace{12mm}

 \centerline{\large Yong-Wan Kim$^{a}$, Hyung Won Lee$^{b}$, and Yun Soo Myung$^{c}$}

\vspace{10mm} {\em Institute of Basic Science and School of
Computer Aided Science \\ Inje University, Gimhae 621-749, Korea}
\vskip .6cm
\end{center}

\begin{center}

\underline{Abstract}
\end{center}

  We  study the propagation of a scalar, the trace of $h_{ij}$ in the deformed Ho\v{r}ava-Lifshitz gravity
   with coupling constant $\lambda$. It turns out that this scalar
   is not a  propagating mode in the Minkowski
   spacetime background. In this work, we do not choose a
   gauge-fixing to identify  the physical degrees of freedom and instead, make it possible  by substituting the constraints
   into the quadratic Lagrangian.

\vspace{15pt}
\baselineskip=18pt
\noindent $^a$ywkim65@gmail.com \\
\noindent $^b$hwlee@inje.ac.kr\\
\noindent $^c$ysmyung@inje.ac.kr

\thispagestyle{empty}

\newpage
\section{Introduction}
Recently Ho\v{r}ava has proposed a renormalizable theory of
gravity at a Lifshitz point\cite{ho1},  which  may be regarded as
a UV complete candidate for general relativity.  Very recently,
the Ho\v{r}ava-Lifshitz gravity theory has been intensively
investigated
in~\cite{ho2,ho3,VW,klu,Nik,Nas,Iza,Vol,CH,CHZ,Nis,KS,OR,Kon,CNPS,SVW},
 its cosmological applications in
~\cite{cos1,cos2}, and its black hole solutions in
~\cite{bh1,bh2}.

We would like to mention that the IR vacuum of this theory is anti
de Sitter (AdS) spacetimes. Hence, it is interesting to take a
limit of the theory, which may lead to  a Minkowski vacuum in the
IR sector. To this end, one may modify the theory by introducing
``$\mu^4R$" and then, take the $\Lambda_W \to 0$ limit. This does
not alter the UV properties of the theory, but it changes the IR
properties. That is, there exists a Minkowski vacuum, instead of
an AdS vacuum.

A relevant issue of (deformed) Ho\v{r}ava-Lifshitz gravity is to
answer to the question of whether it can
 accommodate a scalar mode $\psi \propto H$, the trace of $h_{ij}$,
  in addition to two degrees of freedom for a massless graviton.
Known results were  sensitive to a gauge-fixing. If one chooses a
gauge of $n_i=0$ together with  a Lagrange multiplier $A$, then
there remains a term of $\dot{H}^2$ in the quadratic action, which
may imply that $H$ is physical, but nonpropagating on the
Minkowski background~\cite{ho1,KS}.  On the other hand, choosing a
gauge of $A=E=0$ with a non-dynamical field $B$  leads to two
terms of $c_1\dot{\psi}^2+c_2(\partial_k\psi)^2$, which implies
that a gauge-invariant scalar $\psi$ is a dynamically scalar
degree of freedom~\cite{CHZ}. If the trace $\psi$ is really
propagating on the Minkowski background, the deformed
Ho\v{r}ava-Lifshitz gravity amounts to a scalar-tensor theory.
However, it was known that a choice of gauge-fixing cannot be
done, in general, by substituting the gauge condition into the
action directly\footnote{ In order to find propagators, first
substituting the gauge-condition into the gauge-invariant bilinear
action with parameter $b^2$,  inverting, and then, taking the
limit of $b^2\to \infty$. See Ref.\cite{FPp} for the
gauge-propagator in the Yang-Mills theory, Ref.\cite{HV} for
graviton-propagator in general relativity, Ref.\cite{Stelle} for
graviton-propagator in higher-derivative quantum gravity, and
Ref.\cite{RSS} for graviton-propagator in the Kaluza-Klein
theory.}~\cite{FPp,HV,Stelle,RSS}. Hence, we need to introduce
another approach to confirm the propagation of scalar mode around
Minkowski spacetimes.

In this work, we will not choose any gauge to identify physical
scalar degrees of freedom. One way to identify the physical
degrees of freedom is to treat non-dynamical fields in the
quadratic Lagrangian without fixing a gauge~\cite{FJ,RT}. In this
work, we consider the Lagrangian formalism~\cite{RT} only because
the Hamiltonian formalism was not working for Ho\v{r}ava-Lifshitz
gravity well, and thus, it has shown unwanted results for scalar
degrees of freedom~\cite{LPa}. We would like to mention that there
are two kinds of non-dynamical fields: at the level of quadratic
action, a non-dynamical field may enter the action either linearly
or quadratically.  As is shown in Eq.(\ref{giaction}), for
$\lambda\not=1$, examples of the latter are two gauge-invariant
modes $w_i$ and $\Pi$. These modes can be integrated out:  their
equations can be used to express these in terms of dynamical
fields $\psi$ (the latter enters the action with time derivative)
and then, one gets rid of these  by plugging the resulting
expression
 back into the action. Therefore, the number of dynamical fields
 is not reduced in this way.
 The other is that the action does not contain a quadratic term as a non-dynamical field.
  $A$ is the case for Ho\v{r}ava-Lifshitz gravity
 and $\Phi$ for general relativity. Unlike in the quadratic case,
 the corresponding  equation is a constraint imposed on
 dynamical fields, and thus $A$ is a
 {\it Lagrange multiplier}. An important feature is that the
 constraint reduces the number of dynamical fields.
   This implies that Lagrange multipliers play the important role in
   finding physical degrees of freedom.

   In the view of Faddeev-Jackiw
constraints~\cite{FJ,Jac}, quadratic non-dynamical fields are
superficial constraints and  a linear non-dynamical field is a
true constraint. Hence we wish to distinguish the former with
notation (=) from  the latter with ($\approx$).

In order to compare the foliation-preserving diffeomorphism
(FDiff) of the Ho\v{r}ava-Lifshitz gravity with others, we
introduce transverse diffeomorphism (TDiff), full diffeomorphism
(Diff), and Weyl-transverse diffeomorphism (WTDiff) for general
relativity in the Appendix.

\section{Deformed Ho\v{r}ava-Lifshitz gravity}
First of all, we introduce the ADM formalism where the metric is
parameterized
%%%
\be ds_{ADM}^2= - N^2  dt^2 + g_{ij} \Big(dx^i - N^i dt\Big)
\Big(dx^j - N^j dt\Big)\,, \ee
%%%%
Then, the Einstein-Hilbert action can be expressed as
%%%
\be \label{Eins} S^{EH} = \fft{1}{16\pi G} \int d^4x \sqrt{g} N
\Big(K_{ij} K^{ij} - K^2 + R - 2\Lambda\Big)\,, \ee
%%%%
where $G$ is Newton's constant and extrinsic curvature $K_{ij}$
takes the form
%%%
\be K_{ij} = \fft{1}{2N} \Big(\dot g_{ij} - \nabla_i N_j -
\nabla_j N_i\Big)\,. \ee
%%%%
Here, a dot denotes a derivative with respect to $t$ (
$``~\dot{}~"=\frac{\partial}{\partial t}$).

On the other hand, a deformed action of the non-relativistic
renormalizable gravitational theory  is given by~\cite{KS}
\bea%
\label{act}S^{dHL}&=&\int dtd^3\bx\, \Big({\cal L}_0 +\mu^4R + {\cal L}_1\Big)\,,\\
{\cal L}_0 &=& \sqrt{g}N\left\{\frac{2}{\kappa^2}(K_{ij}K^{ij}
\label{action1}-\lambda K^2)+\frac{\kappa^2\mu^4(\Lambda_W R
  -3\Lambda_W^2)}{8(1-3\lambda)}\right\}\,,\\ {\cal L}_1&=&
\sqrt{g}N\left\{\frac{\kappa^2\mu^4
(1-4\lambda)}{32(1-3\lambda)}R^2 -\frac{\kappa^2}{2w^4}
\left(C_{ij} -\frac{\mu w^2}{2}R_{ij}\right) \left(C^{ij}
-\frac{\mu w^2}{2}R^{ij}\right) \right\}\,.\label{action2}
\eea%
where $C_{ij}$ is the Cotton tensor
%%%
\be C^{ij}=\epsilon^{ik\ell}\nabla_k\left(R^j{}_\ell
-\frac14R\delta_\ell^j\right).\label{def.K.C} \ee
%%%%
Comparing ${\cal L}_0$ with Eq.(\ref{Eins}) of general relativity,
the speed of light, Newton's constant and the cosmological
constant are given by
%%%%
\be c=\fft{\kappa^2\mu}{4}
\sqrt{\fft{\Lambda_W}{1-3\lambda}}\,,\qquad
G=\fft{\kappa^2}{32\pi\,c}\,,\qquad \Lambda=\ft32
\Lambda_W\,.\label{cg} \ee
%%%
The equations of motion were derived in \cite{cos1} and
\cite{bh1}, but we do not write  them  due to the length.

In the limit of $\Lambda_W \to 0$, we obtain  the
$\lambda$-Einstein action from ${\cal L}_0+\mu^4 R$ as
 \be
S^{EH\lambda}=\int dt d^3x \sqrt{g}
N\Bigg[\frac{2}{\kappa^2}\Big(K_{ij}K^{ij}-\lambda K^2\Big)+\mu^4
R\Bigg]\ . \label{SM2} \ee In this case, we have Minkowski
background with ~\cite{KS} \be c^2=\fft{\kappa^2\mu^4}{2}\,,\qquad
G=\fft{\kappa^2}{32\pi\,c}\,,\qquad \Lambda=0\,.\label{mcg} \ee
Considering the $z=3$ Ho\v{r}ava-Lifshitz gravity, we have scaling
dimensions of $[t]=-3,[x]=-1, [\kappa]=0,$ and $[\mu]=1$. We wish
to consider perturbations of the metric around Minkowski
spacetimes, which is a solution of the full theory (\ref{act})
 \be \label{decom1}
g_{ij}= \delta_{ij}+w h_{ij},~N= 1+ wn,~N_i= w n_i. \ee At quadratic
order the action (\ref{SM2}) turns out to be \bea \label{EHlambda}
S^{EH\lambda}_2 &=& w^2\int dt d^3x \Bigg\{{1 \over \kappa^2}
\left[{1\over 2} \dot h_{ij}^2 -{\lambda\over 2} \dot h^2 +
(\partial_i n_j)^2 +(1-2\lambda) (\partial \cdot n)^2 - 2
\partial_i n_j(\dot h_{ij} -\lambda \dot h \delta_{ij})\right]
\nonumber\\
&&\phantom{x} +  {\mu^4\over 2} \left[ -\frac{1}{2}(\partial_k
h_{ij})^2+\frac{1}{2}(\partial_i h)^2 +(\partial_i
h_{ij})^2-\partial_i h_{ij}\partial_j h + 2 n (\partial_i
\partial_j h_{ij}-\partial^2 h) \right]\Bigg\} . \eea

In order to analyze the physical degrees of freedom completely, it
is  convenient to use the cosmological decomposition in terms of
scalar, vector, and tensor modes under spatial  rotations
$SO(3)$~\cite{MFB}
 \bea \label{pert}
 n &=&-\frac{1}{2}A,\nn \\
 n_i&=&\Big(\partial_iB+V_i\Big),\label{decom2} \\
 h_{ij}&=&\Big(\psi\delta_{ij}+\partial_i\partial_j E+2\partial_{(i}F_{j)}+t_{ij}\Big), \nn \eea
where $\partial^iF_i=\partial^iV_i=\partial^it_{ij}=t^i~_i=0$.
 The
last two conditions mean that $t_{ij}$ is a transverse and
traceless tensor in three dimensions.  Using this decomposition,
the scalar modes ($A,B,\psi,E$), the vector modes ($V_i,F_i$), and
the tensor modes ($t_{ij}$) decouple from each other. These all
amount to 10 degrees of freedom for a symmetric tensor in four
dimensions.

Before proceeding, let us check dimensions. We observe
$[n]=0,[n_i]=2,$ and $[h_{ij}]=0$,  which imply
$[A]=0,[B]=1,[V_i]=2,[\psi]=0,[E]=-2,[F_i]=-1,$ and $[t_{ij}]=0$.

The Lagrangian is obtained by substituting (\ref{decom2}) into the
quadratic action (\ref{EHlambda}) as
 \bea \label{fehl}  S^{EH\lambda}_2 &=&
 \int dtd^3x \left\{
\frac{w^2}{2\kappa^2}  \left[  3(1-3\lambda)\dot{\psi}^2
+2\partial_i\omega_j\partial^i\omega^j
-4\left((1-3\lambda)\dot{\psi}+(1-\lambda)\partial^2\dot{E}\right)\partial^2B\right.
\right. \nonumber\\
&& \left.~~~~~~~~~~~~~
+4(1-\lambda)(\partial^2B)^2+2(1-3\lambda)\dot{\psi}\partial^2\dot{E}
+(1-\lambda)(\partial^2\dot{E})^2+ \dot{t}_{ij}\dot{t^{ij}}\right]
\nonumber\\
&&\left.~~~~~~~~~~~~~
+\frac{\mu^4w^2}{4}\left[2\partial_k\psi\partial^k\psi
+4A\partial^2\psi -\partial_k t_{ij}\partial^k
t^{ij}\right]\right\} \eea with $w_i=V_i-\dot{F}_i$.

On the other hand, the higher order action obtained from ${\cal
L}_1$ takes the form  \bea S^1_2=\int dt d^3x
\frac{\kappa^2\mu^2w^2}{8}\Bigg\{
&-&\frac{1+\lambda}{2(1-3\lambda)} \psi
\partial^4 \psi -\frac{1}{4}t_{ij}\partial^4 t^{ij} \nn \\
&+&\frac{1}{\mu w^2} \epsilon^{ijk} t_{il} \partial^4
\partial_j t^l~_k+\frac{1}{\mu^2w^4}
t_{ij} \partial^6 t^{ij}
 \Bigg\}. \label{1action} \eea
We observe that  two modes of $\psi$ and $t_{ij}$ exist in the
higher order action.

 Now we are in a
position to discuss the diffeomorphism in the $z=3$
Ho\v{r}ava-Lifshitz gravity. Since the anisotropic scaling of
temporal and spatial coordinates ($t\to b^z t, x^i \to b x^i$),
the time coordinate $t$ plays a privileged role. Hence, the
spacetime symmetry is smaller than the full diffeomorphism (Diff)
in the standard general relativity (Einstein gravity).  The
Ho\v{r}ava-Lifshitz gravity of $S^{EH\lambda}_2+S^1_2$ should be
invariant under the ``foliation-preserving" diffeomorphism (FDiff)
whose form is given by \be t \to \tilde{t}=t+\epsilon^0(t),~~x^i
\to \tilde{x}^i=x^i+\epsilon^i(t,\bf{x}). \ee Using the notation
of $\epsilon^\mu=(\epsilon^0,\epsilon^i)$ and
$\epsilon_\nu=\eta_{\nu\mu}\epsilon^\mu$, the perturbation of
metric transforms as \be \delta g_{\mu\nu} \to
\delta\tilde{g}_{\mu\nu}=\delta g_{\mu\nu}+\partial_\mu
\epsilon_\nu+\partial_\nu \epsilon_\mu. \ee Further, making a
decomposition $\epsilon^i$ into a scalar $\xi$ and a pure vector
$\zeta^i$ as $\epsilon^i=\partial^i\xi+ \zeta^i$ with $\partial_i
\zeta^i=0$, one finds the transformation for the scalars \be
\label{trans1} A \to \tilde{A}=A-2\dot{\epsilon^0},~\psi \to
\tilde{\psi}=\psi,~ B \to \tilde{B}=B+\dot{\xi},~E \to
\tilde{E}=E+2\xi.\ee On the other hand, the vector and the tensor
take the forms \be \label{trans2} V_i \to \tilde{V}_i=V_i
+\dot{\zeta}_i,~F_i\to \tilde{F}_i=F_i+\zeta_i,~t_{ij} \to
\tilde{t}_{ij}=t_{ij}. \ee Considering scaling dimensions of
$[\epsilon^0]=-3$ and $[\epsilon^i]=-1$,  we have $[\xi]=-2$ and
$[\zeta^i]=-1$. For the FDiff transformations,  gauge-invariant
combinations are \be t_{ij},~~w_i=V_i-\dot{F}_i, \ee for tensor
and vector, respectively and \be \Big(\psi,~~\Pi=2B-\dot{E} \Big)
\ee for two  scalar modes. Finally, we note scaling dimensions:
$[w_i]=2$ and $[\Pi]=1$. We emphasize that ``$A$" leaves a
gauge-dependent quantity alone. For other gauge-invariant scalars
in  general relativity, see the Appendix.

\section{$n_i=0$ gauge-fixing}

Firstly, we may consider a gauge of $n_i=0$~\cite{ho1,KS}. It
amounts to  the gauge-fixing: \be B=0,~V_i=0.\ee Then, the
bilinear action takes the form
 \bea \label{nizact}  S^{EH\lambda}_2 =
 \int dtd^3x \Bigg\{
\frac{\omega^2}{2\kappa^2}  \Big[  3(1-3\lambda)\dot{\psi}^2
&+&2\partial_i\dot{F}_j\partial^i\dot{F}^j
+2(1-3\lambda)\dot{\psi}\partial^2\dot{E}
+(1-\lambda)(\partial^2\dot{E})^2+ \dot{t}_{ij}\dot{t^{ij}}\Big]
\nonumber\\
&+&\frac{\mu^4\omega^2}{4}\left[2\partial_k\psi\partial^k\psi
+4A\partial^2\psi -\partial_k t_{ij}\partial^k
t^{ij}\right]\Bigg\}. \eea It is obvious that $A$ is a Lagrange
multiplier and thus it provides a constraint \be
\label{aconstraint}
\partial^2\psi\approx0.\ee
It is emphasized that  the notation ``$\approx$" is used to denote
the constraint obtained by varying the Lagrange multiplier only.
We may consider $\dot{E}$ and $\dot{F}_i$ as non-dynamical fields
even though they have time derivatives. Since gauge-invariant
quantities are given by $\Pi=2B-\dot{E}$ and $w_i=V_i-\dot{F}_i$,
it seems that canonical variables are not $E$ and $F_i$ but
$\dot{E}$ and $\dot{F}_i$.
 Hence,
in order to eliminate these fields, we use their variations \be
\partial^2 \dot{E}=-\Big(\frac{1-3\lambda}{1-\lambda}\Big)\dot{\psi},~~\dot{F}_i=0. \ee
Substituting these into the quadratic action, we have the relevant
one
 \be   \label{npert} S^{EH\lambda}_2=\int dt d^3x\Bigg\{
\frac{w^2}{2\kappa^2}\Bigg(
\frac{2(1-3\lambda)}{1-\lambda}\dot{\psi}^2+\dot{t}_{ij}\dot{t}^{ij}
\Bigg) -\frac{\mu^4w^2}{4}
\partial_k t_{ij} \partial^k t^{ij}
 \Bigg\}. \ee

It is clear that for $\lambda\not=1,1/3$, the scalar ``$\psi$" is
not a propagating mode on  the Minkowski background because of the
constraint (\ref{aconstraint}), while $t_{ij}$ represents for a
massless graviton propagation. On the other hand, the bilinear
action to ${\cal L}_1$ leads to
 \be S^1_2=\int dt d^3x
\frac{\kappa^2\mu^2w^2}{8}\Bigg\{ -\frac{1}{4}t_{ij}\partial^4
t^{ij} +\frac{1}{\mu w^2} \epsilon^{ijk} t_{il} \partial^4
\partial_j t^l~_k+\frac{1}{\mu^2w^4}
t_{ij} \partial^6 t^{ij}
 \Bigg\}. \label{1actn} \ee
Plugging \be \psi \to \frac{1-\lambda}{2(1-3\lambda)}H,~~t_{ij}\to
\tilde{H}_{ij}\ee into Eqs. (\ref{npert}) and (\ref{1actn}) with
$x^0=ct ~([x_0]=-1,[c]=2)$, one arrives at the quadratic action
exactly~\cite{KS}
 \bea &&S^{HL}_2=\int dx^0 d^3x
\Bigg\{\frac{w^2c}{2\kappa^2}\left[\Big(\partial_0\tilde{H}_{ij}\Big)^2-\frac{\mu^4\kappa^2}{2c^2
} \Big(\partial_k \tilde{H}_{ij}\Big)^2\right]
+\frac{w^2c(1-\lambda)}{4\kappa^2(1-3\lambda)}\Big(\partial_0{H}\Big)^2
\nn \\
&&+ \frac{\kappa^2 \mu^2 w^2}{8c}\Bigg[
-\frac{1}{4}\tilde{H}_{ij}\partial^4 \tilde{H}^{ij} +\frac{1}{\mu
w^2} \epsilon^{ijk} \tilde{H}_{il}
\partial^4
\partial_j \tilde{H}^l~_k+\frac{1}{\mu^2w^4}
\tilde{H}_{ij} \partial^6 \tilde{H}^{ij} \Bigg]\Bigg\}.
\label{senn} \eea  Note that for $1/3<\lambda<1$, the kinetic term
of $H$ becomes negative, indicating a ghost instability. Thus, one
may argue that either $\lambda$ runs to $1^+$ from above in the IR
or $H$ does not couple at all to matter. However, this may  not be
a promising  way to resolve the ghost problem. A correct answer is
that the scalar mode of $H \propto \psi$ is a nonpropagating mode.

We also see  from (\ref{senn})  that the speed of gravitational
interaction is \be c_g^2=\frac{\mu^4\kappa^2}{2c^2 }c_0^2, \ee
where $c_0^2$ is the speed of light. We know that the propagation
of gravity interaction equals the velocity of light to better than
$1:1000$. Hence, we get that \be c^2=\frac{\mu^4\kappa^2}{2 } \ee
with the above accuracy, independent of the value of the
couplings.

Finally, we have the quadratic action \bea &&S^{HL}_2=\int dx^0
d^3x
\Bigg\{\frac{w^2c}{2\kappa^2}\Big(\partial_\mu\tilde{H}_{ij}\Big)^2
+\frac{w^2c(1-\lambda)}{4\kappa^2(1-3\lambda)}\Big(\partial_0{H}\Big)^2
\nn \\
&&+ \frac{\kappa^2 \mu^2 w^2}{8c}\Bigg[
-\frac{1}{4}\tilde{H}_{ij}\partial^4 \tilde{H}^{ij} +\frac{1}{\mu
w^2} \epsilon^{ijk} \tilde{H}_{il}
\partial^4
\partial_j \tilde{H}^l~_k+\frac{1}{\mu^2w^4}
\tilde{H}_{ij} \partial^6 \tilde{H}^{ij} \Bigg]\Bigg\}.
\label{senf} \eea

\section{$A=0$ and $E=0$ gauge-fixing }

In the perturbation, the lapse function $n$ is  a function of $t$
only, and thus, $A$ is  a function of $t$. It may allow $A$ to be
a gauge degree of freedom  by choosing a initial time $t_0$. Also,
we may choose $E$ as a gauge degree of freedom. In this section,
we start with  a gauge-fixing ~\cite{CHZ}: \be A=0,~E=0.\ee Then,
the bilinear action takes the form
 \bea S^{EH\lambda}_2=\int dt d^3x\Bigg\{
\frac{w^2}{2\kappa^2}&\Bigg(&
3(1-3\lambda)\dot{\psi}^2-4(1-3\lambda)\dot{\psi}\partial^2
B+4(1-\lambda)B \partial^4 B \label{qir1} \\
+2\partial_k w_i \partial^k w^i&+&\dot{t}_{ij}\dot{t}^{ij} \Bigg)
+\frac{\mu^4w^2}{4}\Bigg( 2\partial_k\psi
\partial^k \psi -
\partial_k t_{ij} \partial^k t^{ij}\Bigg)
 \Bigg\}.
 \label{qir2}
\eea For $\psi \to -2 \Psi$, the first line (\ref{qir1}) recovers
those of Ref.\cite{CHZ}  with $a=1$ and $\Lambda=0$. We observe
that $B$ and $w_i$ are non-dynamical fields  because they do not
have time derivatives. Hence, in order to eliminate these, we use
their variations \be
\partial^2 B=\frac{(1-3\lambda)}{2(1-\lambda)}\dot{\psi},~~w_i=0. \ee
Substituting these relations  into the quadratic action, we have
the relevant one
 \be S^{EH\lambda}_2=\int dt d^3x\Bigg\{
\frac{w^2}{2\kappa^2}\Bigg(
\frac{2(1-3\lambda)}{1-\lambda}\dot{\psi}^2+\dot{t}_{ij}\dot{t}^{ij}
\Bigg) +\frac{\mu^4w^2}{4}\Bigg( 2(\partial_k\psi)^2  -
\partial_k t_{ij} \partial^k t^{ij}\Bigg)
 \Bigg\}. \ee
 \label{physact}
It seems that for $\lambda\not=1,1/3$, the scalar ``$\psi$" is
propagating  on the Minkowski background, in addition to $t_{ij}$
for a massless graviton propagation. This  is because a  kinetic
term $(\partial_k\psi)^2$ survives because a gauge condition of
$A=0$ does not impose any constraint. However, the sign of
$(\partial_k\psi)^2$ is opposite to that of $\partial_k t_{ij}
\partial^k t^{ij}$ and thus, it may not lead to a proper
scalar propagation on the Minkowski background.

On the other hand, the bilinear action to ${\cal L}_1$ leads to
 \bea S^1_2=\int dt d^3x
\frac{\kappa^2\mu^2w^2}{8}\Bigg\{
&-&\frac{1+\lambda}{2(1-3\lambda)} \psi
\partial^4 \psi -\frac{1}{4}t_{ij}\partial^4 t^{ij} \nn \\
&+&\frac{1}{\mu w^2} \epsilon^{ijk} t_{il} \partial^4
\partial_j t^l~_k+\frac{1}{\mu^2w^4}
t_{ij} \partial^6 t^{ij}
 \Bigg\}, \label{1act} \eea
where the first term represents a fourth order for the scalar
$\psi$. This term survives because a gauge condition of $A=0$ was
chosen.

\section{Without gauge-fixing}
One may identify physical degrees of freedom, without fixing any
gauge, by treating non-dynamical fields in (\ref{fehl})properly.
First of all, we express the quadratic action (\ref{fehl}) in
terms of gauge-invariant quantities of  the scalar, vector, and
tensor modes as \bea S^{EH\lambda}_2 =
 \int dtd^3x &\Bigg\{&
\frac{w^2}{2\kappa^2}  \Big[  3(1-3\lambda)\dot{\psi}^2
-2w_i\bigtriangleup\omega^i
-2(1-3\lambda)\dot{\psi}\bigtriangleup \Pi+(1-\lambda)(\bigtriangleup\Pi)^2 \nonumber \\
\label{giaction}&+& \dot{t}_{ij}\dot{t}^{ij}\Big]
+\frac{\mu^4w^2}{4}\left[-2\psi\bigtriangleup\psi
+4A\bigtriangleup\psi + t_{ij} \bigtriangleup t^{ij}\right]\Bigg\}
\eea with $\bigtriangleup=\partial_i\partial^i=\partial^2$. We
note that $S_2^1$ in (\ref{1action}) contains only $\psi$ and
$t_{ij}$, which are also gauge-invariant. It is emphasized again
that ``$A$" is not a gauge-invariant quantity and thus, it should
be eliminated in the consistent quadratic action. Fortunately,
this is possible because it belongs to a Lagrange multiplier,
irrespective of any value $\lambda$.

Before proceeding, we mention two special cases: $\lambda=1/3$ and
$\lambda=1$. Plugging $\lambda=1/3$ into the above action, we have
a term like $\dot{\psi}^2$. In addition, we have two non-dynamical
fields ($w_i,~\Pi$) and one Lagrange multiplier ($A$) which
provide two relations and one constraint as, respectively\be
\bigtriangleup w_i=0,~~\bigtriangleup
\Pi=0,~~\bigtriangleup\psi\approx 0. \ee This implies that the
$t_{ij}$ are only  propagating tensor modes. Similarly,  for
$\lambda=1$, one  have
 no scalar mode $\psi$ definitely because of one relation and two constraints from
 one non-dynamical field ($w_i$) and two Lagrange multipliers
 ($\Pi,~A$):
  \be
  \bigtriangleup w_i=0,~\dot{\psi}\approx 0,~ \bigtriangleup
  \psi\approx0.\ee

Note here that for $\lambda\not=1/3,1$, $\Pi$ and $w_i$ are  two
non-dynamical fields  to be solved  to have two relations \be
\bigtriangleup
\Pi=\frac{(1-3\lambda)}{(1-\lambda)}\dot{\psi},~~w_i=0. \ee
Substituting these into the quadratic action, we have \bea
S^{EH\lambda}_2 &=&
 \int dtd^3x \left\{
\frac{\omega^2}{2\kappa^2}  \left[
\frac{2(1-3\lambda)}{(1-\lambda)}\dot{\psi}^2
+\dot{t}_{ij}\dot{t^{ij}}\right]\right.\nonumber\\
&&~~~~~~~~~~\left.+\frac{\mu^4\omega^2}{4}\left[-2\psi\bigtriangleup\psi
+4A\bigtriangleup \psi + t_{ij}\bigtriangleup t^{ij}\right]
\right\}.\eea Here we observe that for $1/3<\lambda<1$, a ghost
appears because there is a negative kinetic term for $\psi$. Also,
comparing $-2\psi\bigtriangleup\psi$ with $t_{ij}\bigtriangleup
t^{ij}$, we find a negative spatial derivative term for scalar
$\psi$. Hence it should not  be a propagating mode on the
Minkowski background.  Since $A$ is  a Lagrange multiplier, its
variation provides a constraint \be \bigtriangleup\psi\approx 0.
\ee Then, we have the bilinear action without $A$

\be S^{EH\lambda}_2 = \frac{\omega^2c}{2\kappa^2}
 \int d^4x
  \left[
\frac{2(1-3\lambda)}{(1-\lambda)}(\partial_0\psi)^2
+\partial_0{t}_{ij}\partial_0{t^{ij}} -\frac{\mu^4\kappa^2}{2c^2}
\partial_k t_{ij}
\partial^k t^{ij}\right]\ee
with $x^0=ct$. Using $c^2=\mu^4 \kappa^2/2$, we may have the
relativistic action for graviton \be S^{EH\lambda}_2 =
\frac{\omega^2c}{2\kappa^2}
 \int d^4x
  \left[
\frac{2(1-3\lambda)}{(1-\lambda)}(\partial_0\psi)^2 +{t}_{ij}
\square t^{ij} \right],\ee
 which implies that the scalar mode is not
propagating even for $\lambda>1$ because it contains
$(\partial_0\psi)^2$ only, while the tensor mode (graviton) is
propagating on the Minkowski background. Here
$\square=\eta^{\mu\nu}\partial_\mu\partial_\nu$ with
$\eta_{\mu\nu}={\rm diag}(-,+,+,+)$. Finally, the higher order
action $S^1_2$ is given by (\ref{npert}). However, this action
does not determine whether  a mode is propagating or not.

We would like to mention that $\psi$ is a non-propagating mode
under the $n_i=0$ gauge with a Lagrange multiplier $A$ in Section
3, but it is a propagating mode  under the  $A=E=0$ gauge with two
non-dynamical fields $B$ and $w_i$ in Section 4. It seems that the
origin of this discrepancy is due to different gauge-fixings.
However, it was known that a choice of gauge-fixing cannot be
done, in general, by substituting the gauge condition into the
action directly~\cite{FPp,HV,Stelle,RSS}. Hence, our approach is a
consistent mathematical formalism  for checking the absence of new
degrees of freedom around the Minkowski background.

\section{Discussions}

A hot issue of Ho\v{r}ava-Lifshitz gravity is to clarify  whether
it can
 accommodate a scalar mode as the trace of $h_{ij}$, in addition to
  two degrees of freedom for a massless graviton.
Actually, known results were  sensitive to a gauge-fixing. If one
chooses a gauge of $n_i=0 ~(B=V_i=0)$ together with  a Lagrange
multiplier $A$ (equivalently, $\partial^2\psi\approx0$) and two
non-dynamical fields ($\dot{E},\dot{F}_i$) there remains a term of
$\dot{H}^2$ in the action, which  implies that $H$ is
nonpropagating~\cite{ho1,KS}. On the other hand, choosing a gauge
of $A=E=0$ together with two non-dynamical fields ($B,w_i$) leads
to two terms of $c_1\dot{\psi}^2+c_2(\partial_k\psi)^2$, which may
imply that a gauge-invariant scalar $\psi$ is a propagating scalar
degree of freedom~\cite{CHZ}.

In this work, we did not choose any gauge to identify physical
scalar degrees of freedom.  Without fixing a gauge, one could
identify physical degrees of freedom  by treating two
non-dynamical fields  ($w_i,\Pi,$) and one Lagrange multiplier $
A$ appropriately.  This means that Lagrange multiplier plays the
important role in finding physical degrees of freedom.  In the
foliation-preserving diffeomorphism (FDiff), gauge-invariant
scalars are $\psi$ and $\Pi$, while the lapse
perturbation``$A\propto n$" is a gauge-dependent scalar. Thus, the
latter should be eliminated from the quadratic action. It is
either a function $A(t)$ when imposing the projectability
condition or a function $A(t,{\bf x})$ without the projectability
condition.  Because $A$ is a Lagrange multiplier, we could always
use it to obtain a constraint $\bigtriangleup\psi\approx 0$ and
thus, $\psi$ is not a propagating scalar mode. A gauge-invariant
scalar $\Phi=A-\dot{\Pi}$ emerging  in general relativity is split
into a gauge-dependent scalar $A$ and a gauge-invariant scalar
$\Pi$, due to the FDiff. We note that $\Phi$ is a Lagrange
multiplier in  TDiff and Diff as well as $A$ is  Lagrange
multipliers in the deformed Ho\v{r}ava-Lifshitz gravity.

 We compare FDiff with different diffeomorphisms in general
relativity in Table 1.  As  the general analysis was  shown  in
the Appendix, it is not easy to have a scalar mode in
four-dimensional general relativity. The TDiff case has less
symmetry  than Diff and WTDiff cases. One  has to realize that the
TDiff case has three gauge-invariant scalars, thanks to an
additional condition of $\partial_\mu \epsilon^\mu=0$. This case
provides really  a scalar mode which is propagating on the
Minkowski background. Two cases of Diff and WTDiff correspond to
enhanced diffeomorphisms. As a result, there are  two
gauge-invariant scalars and thus, no propagating scalar mode. The
FDiff of the Ho\v{r}ava-Lifshitz gravity is similar to Diff and
WTDiff cases, which have  enhanced diffeomorphisms,  compared with
the TDiff. Hence, we expect to have no propagating scalar mode in
the deformed Ho\v{r}ava-Lifshitz gravity.

We would like to mention a couple of recent works. The
authors~\cite{CNPS} have shown that $\psi$ is a scalar degree of
freedom appeared when the massless limit of a massive graviton
(vDVZ discontinuity~\cite{vDVZ}).  Using the Hamiltonian
constraints, the authors~\cite{SVW} have argued  that a scalar
mode of $\psi$ is propagating around the Minkowski space but it
has a negative kinetic term, providing a ghost mode~\cite{SVW}.
Hence, it was strongly suggested that it is desirable to eliminate
this scalar mode if at all possible.

Consequently, we have shown that the  deformed Ho\v{r}ava-Lifshitz
gravity has no scalar mode which is propagating on the Minkowski
background.

\begin{table}
 \caption{Summary for scalar modes. GR (HL) means general relativity (deformed Ho\v{r}ava-Lifshitz gravity).
 GIS denotes gauge-invariant scalars.  SDoF and TDoF mean number of scalar and tensor
 degrees of freedom, respectively. Here $\Phi=A-2\dot{B}+\ddot{E}=A-\dot{\Pi}$, $\Theta=A-\bigtriangleup E$, and
 $\Pi=2B-\dot{E}$. }
\begin{tabular}{|c|c|c|c|c|}
  \hline
 diffeomorphism  & TDiff & Diff & WTDiff &FDiff \\
  \hline
 Theory & GR & GR & GR & HL \\ \hline
 parameters & $a\not=1,b\not=1$ & $a=b=1$ & $a=1/2,b=3/8$& $\lambda\not=1,1/3$ \\ \hline
 GIS& $\psi,\Phi,\Theta$  & $\psi,\Phi$ & $\Xi=\psi+\Phi,\Upsilon=\psi+\Theta$& $\psi,\Pi$ \\ \hline
  SDoF & 1($\psi$) & 0 &0 & 0 \\ \hline
   TDoF & 2($t_{ij}$) &  2($t_{ij}$) &  2($t_{ij}$)&  2($t_{ij}$) \\
  \hline
\end{tabular}
\end{table}

{\it Note added}--after the present work was released, relevant
works on extra scalar mode have  appeared on the arXiv. The
authors~\cite{GWBR} have shown that on the cosmological
background, the extra  scalar is non-dynamical. One of authors has
found that $\psi$ is a scalar degree of freedom related to  the
massless limit of  the case with Fierz-Pauli mass
terms~\cite{myungm}. However, using the Lorentz-violating mass
terms, there is no such a scalar appeared  in the massless limit.
Also, the authors in \cite{BPS} have found that for a general
background, the extra mode is propagating. The extra mode
satisfies equation of motion which is  first order in time
derivatives. At linear level, thus,  the mode is manifest only
around spatially inhomogeneous and time-dependent background with
two serious problems.  However, the Minkowski spacetime is  a
singular point. Furthermore, the authors \cite{BS} have shown that
the extra mode is not allowed  because of its ghost-like
instability around the Minkowski background.

\section*{Acknowledgement}

Y. Kim was supported by the Korea Research Foundation Grant funded
by Korea Government (MOEHRD): KRF-2007-359-C00007. H. Lee was
supported by KOSEF, Astrophysical Research Center for Structure
and Evolution of the Cosmos at Sejong University. Y. S. Myung  was
supported by the Korea Research Foundation (KRF-2006-311-C00249)
funded by the Korea Government (MOEHRD).

\section*{Appendix: General relativity with different diffeomorphisms}
The most general relativistic Lagrangian for a massless symmetric
tensor field $h_{\mu\nu}$ is given by~\cite{ABGV,blas} \be {\cal
L}_{GR}= {\cal L}^I+\beta {\cal L}^{II}+ a {\cal L}^{III}+ b{\cal
L}^{IV},\ee where \bea {\cal L}^I&=&\frac{1}{4}\partial_\mu
h^{\nu\rho}\partial^\mu h_{\nu\rho},~{\cal
L}^{II}=-\frac{1}{2}\partial_\mu h^{\mu\rho}\partial_\nu
h^\nu~_\rho,\nn \\
{\cal L}^{III} &=& \frac{1}{2} \partial^\mu h \partial^\rho
h_{\mu\rho},~~{\cal L}^{IV}=-\frac{1}{4} \partial_\mu h
\partial^\mu h.
\eea Under a general transformation of the fields $h_{\mu\nu} \to
h_{\mu\nu}+\delta h_{\mu\nu}$, we have up to total derivatives
\bea \delta{\cal L}^I&=-&\frac{1}{2}\delta h_{\mu\nu} \square
h^{\mu\nu},~\delta{\cal L}^{II}=\delta h_{\mu\nu}\partial^\rho
\partial^{(\mu}
h^{\nu)}~_\rho,\nn \\
{\cal L}^{III} &=& -\frac{1}{2} \Big(\delta h
\partial^\mu\partial^\nu h_{\mu\nu}+\delta h_{\mu\nu}\partial^\mu\partial^\nu h\Big),~~{\cal L}^{IV}=\frac{1}{2}
\delta  h \square h. \eea We note that the vector Lagrangian is
problematic unless $\beta=1$ because it induces a ghost
problem~\cite{blas}. Hence, we choose $\beta=1$ case. It follows
that the combination \be \label{tdiff} {\cal L}_{\rm TDiff}={\cal
L}^I+ {\cal L}^{II}+ a {\cal L}^{III}+ b{\cal L}^{IV},\ee with
arbitrary $a$ and $b$ is invariant under restricted gauge
transformations \be \label{gaug1}\delta h_{\mu\nu}=\partial_\mu
\epsilon_\nu+
\partial_\nu \epsilon_\mu \ee with \be \label{gaug2}
\partial_\mu\epsilon^\mu=0. \ee
It is noted  that $\epsilon^0(t,{\bf x})$ and  $\epsilon^i(t,{\bf
x})$. We call the transformations (\ref{gaug1}) and (\ref{gaug2})
transverse diffeomorphisms (TDiff)~\cite{SV,BVN}. We can obtain
two enhanced gauge symmetries by adjusting  parameters $a$ and
$b$: Firstly, $a=b=1$ leads to the Fierz-Pauli Lagrangian which is
invariant under the full diffeomorphisms (Diff), where the
condition (\ref{gaug2}) is dropped~\cite{FP}. This corresponds to
the standard general relativity (Einstein gravity). Secondly,
$a=1/2,b=3/8$ provides  Weyl symmetry of $h_{\mu\nu} \to
h_{\mu\nu}+\frac{\phi}{2} \eta_{\mu\nu}$, in addition to TDiff. We
call this enhanced symmetry the Weyl-transverse diffeomorphisms
(WTDiff)~\cite{ABGV}.

Now let us investigate  mode propagations when using the TDiff.
 Considering the decomposition (\ref{decom1}) with (\ref{decom2}),
  we have the same transformations in Eqs.(\ref{trans1}) and
  (\ref{trans2}) except replacing $B\to
\tilde{B}=B+\dot{\xi} $ by
  \be B\to
\tilde{B}=B-\epsilon^0+\dot{\xi} \ee in general relativity. In
this case, using the residual gauge condition of Eq.(\ref{gaug2})
which implies $\dot{\epsilon}_0=\partial^2\xi$, we have three
gauge-invariant scalars, \be
\Big(\psi,~\Phi=A-2\dot{B}+\ddot{E},~\Theta=A-\partial^2E\Big).
\ee Substituting (\ref{decom1}) and (\ref{decom2}) into
(\ref{tdiff}) leads to \be {\cal L}_{\rm TDiff}= {\cal L}^t_{\rm
TDiff}+{\cal L}^v_{\rm TDiff}+{\cal L}^s_{\rm TDiff}, \ee where
\be {\cal L}^t_{\rm TDiff}=\frac{1}{4} t_{ij} \square
t^{ij},~~{\cal L}^v_{\rm TDiff}=-\frac{1}{2} w_i \bigtriangleup
w^i, \ee for tensor and vector modes and \bea {\cal L}^s_{\rm
TDiff}&=& \frac{1}{4} \Big(3 \dot{\psi}^2+\psi \bigtriangleup \psi
-\dot{\Theta}^2-\Theta \bigtriangleup(\Theta-2\Phi)-2
\bigtriangleup \psi (\Phi-\Theta) \Big)
\nn \\
&+& \frac{a}{2}\Big( (\Theta -3
\psi)(\bigtriangleup(\Theta-\psi-\Phi)-\ddot{\Theta}) \Big) \\
&-&  \frac{b}{4}\Big( (\dot{\Theta} -3 \dot{\psi})^2+ (\Theta
-3\psi)\bigtriangleup(\Theta-3\psi) \Big) \eea  for all scalar
modes. From this decomposition, we realize that $\Phi$ is always a
Lagrange multiplier whose variation yields the constraint \be
\bigtriangleup\Big[(1-3a)\psi-(1-a)\Theta \Big]\approx 0. \ee In
this case, the Lagrangian reduces to \be {\cal L}^s_{\rm
TDiff}=\frac{Z}{(a-1)^2} \psi \square \psi,~~{\rm with}~
Z=\frac{3}{2}\Big(a-\frac{1}{3}\Big)^2-\Big(b-\frac{1}{3}\Big)\ee
which implies that for $b<1/3$, $\psi$ is really  a propagating
scalar mode on the Minkowski background. For two cases of $a=b=1$
and $a=1/2,b=3/8$, we have $Z=0$, which implies that these should
be treated separately.

In the Diff  case of $a=b=1$, only two scalar combinations  are
gauge invariant, namely \be \Big(\Phi,~\psi\Big). \ee Then, its
Lagrangian takes the form \be {\cal L}^s_{\rm Diff}=-\frac{1}{2}
\Big(-2\Phi \bigtriangleup \psi +3\dot{\psi}^2 +\psi
\bigtriangleup \psi\Big).  \ee However, since $\Phi$ is a Lagrange
multiplier, its variation leads to $\bigtriangleup \psi\approx0$.
Plugging this into the above, we have
 \be {\cal L}^s_{\rm Diff}=-\frac{3}{2}
\dot{\psi}^2,  \ee which means that $\psi$ is  not propagating on
the Minkowski background.

Finally, for Weyl transformations of $a=1/2$ and $b=3/8$, we have
two scalar invariants which are also scalar for TDiff, \be
\Big(\Xi=\Phi+\psi,~~\Upsilon=\Theta+\psi\Big). \ee Then, its
Lagrangian is given by \be {\cal L}^s_{\rm
WTDiff}=-\frac{1}{96}\Big(2(8\Xi-3~\Upsilon)\bigtriangleup
\Upsilon-6\dot{\Upsilon}^2\Big). \ee However, since $\Xi$ is a
Lagrange multiplier, its variation leads to $\bigtriangleup
\Upsilon\approx 0$. Plugging this into the above, we have
 \be {\cal L}^s_{\rm WTDiff}=-\frac{1}{16}
\dot{\Upsilon}^2,  \ee which means that $\Upsilon$ is  not
propagating on the Minkowski background.

\end{document}